\documentclass[twocolumn,twoside,letterpaper,11pt]{article}
\usepackage{graphicx,xrrc,natbib}
\usepackage{times}

\begin{document}


\title{AIN'T NO CRAB, PWN GOT A BRAND NEW BAG: CORRELATED RADIO AND X-RAY 
STRUCTURES IN PULSAR WIND NEBULAE}


%
%
%
%


\author{    M. S. E. Roberts                        } 
\institute{ McGill University/Eureka Scientific         } 
\address{ 3600 University St., Montr\'eal, QC, H3A 2T8, Canada          } 
\email{     roberts@physics.mcgill.ca                         } 

\author{    M. Lyutikov, B. M. Gaensler, C. Brogan}
\email{     lyutikov@cita.utoronto.ca, bgaensler@cfa.harvard.edu, cbrogan@ifa.hawaii.edu}

\author{C. R. Tam, R. W. Romani}
\email{tamc@physics.mcgill.ca, rwr@stanford.edu}


\maketitle

\abstract{The traditional view of radio pulsar wind nebulae (PWN), encouraged
by the Crab nebula's X-ray and radio morphologies,  is that they are a 
result of the integrated history of their pulsars' wind. 
The radio emission should therefore be largely unaffected by recent pulsar
activity, and hence minimally correlated with structures in the 
X-ray nebulae. Observations
of several PWN, both stationary (sPWN) and rapidly moving (rPWN),
now show clear morphological relationships between structures in the
radio and X-ray with radio 
intensity variations on the order of unity. 
We present high-resolution X-ray and
radio images of several PWN of both types and discuss the morphological 
relationships between the two wavebands. }

\section{Introduction}

The traditional model of radio pulsar wind nebulae (PWN), encouraged by the 
amorphous shape of the Crab radio nebula, is where the bubble blown by the 
pulsar acts as a bag which is filled with particles and magnetic field 
deposited by the wind since the pulsar's birth \citep[for a recent review of 
PWN, see][]{krh04}. 
The flow velocity of freshly injected material at the termination
shock is $v_0 \sim 0.5$c. As the material expands into the nebula, the 
material will slow, but the nebular crossing time is still on the order of 
tens to a few hundreds of years, much less than the $\sim 1000-10,000$yr 
typical age of the nebula. The synchrotron cooling time of radio 
emitting electrons is probably much longer than the age of the 
nebula, while the cooling time of the X-ray emitting electrons is 
generally thought to be on the order of the crossing time.   
The long-lived radio 
emitting particles therefore reflect the integrated history of the 
pulsar wind output and should be fairly evenly distributed throughout
the nebula and be largely unaffected by recent pulsar 
activity. This is in contrast to the relatively short-lived, 
X-ray emitting particles which should reflect the current pulsar wind 
conditions.
However, high-resolution observations of both stationary (sPWN) and 
rapidly-moving (rPWN) nebulae often show clear morphological relationships
between X-ray and radio structures within the nebulae. Sometimes these
are correlations and sometimes anti-correlations, but it is apparent that
X-ray structures are often associated with radio intensity variations on the
order of unity. These require local magnetic field variations which are 
related to the present day wind activity and can potentially be used 
to probe the structure of the pulsar wind. 

\section{Composite SNR With SPWN}

A pulsar is born in a supernova explosion, and so the 
initial environment of a young PWN is that of the expanding ejecta
of the progenitor star. In this environment, the young PWN 
undergoes accelerated expansion which generally greatly outpaces any
motion of the pulsar imparted by the supernova. Hence the morphology of the
PWN is not greatly effected by the displacement of the pulsar from its
birthsite, and we will refer to such a nebula as a stationary PWN (sPWN).  
If the PWN is surrounded by a radio and/or X-ray shell caused by the supernova 
blast wave, then the system is referred to as a composite supernova remnant 
(SNR) and there are observational indications of the expanding ejecta's
properties. 

\subsection{SNR G11.2$-$0.3}

\begin{figure}
  \begin{center}
    \includegraphics[width=\columnwidth]{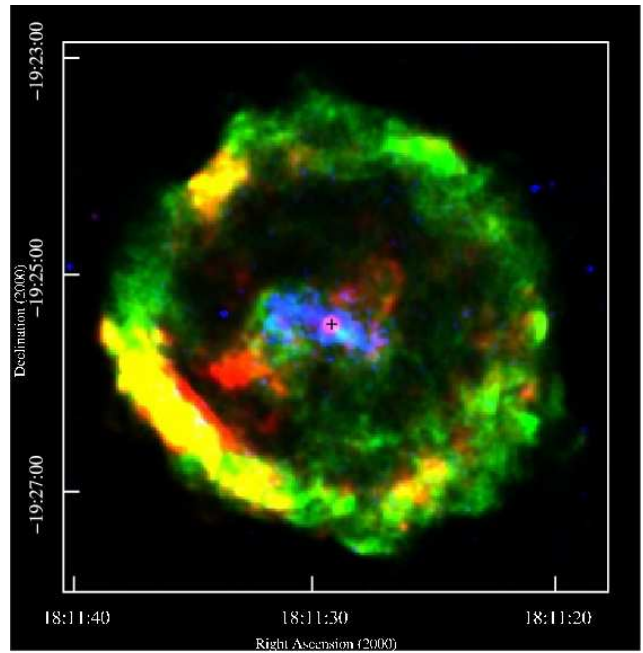}
    \caption{\small {\it Chandra} and VLA image of SNR G11.2$-$0.3. 
Red is thermal X-ray emission, green is 3.5~cm radio emission, and blue is 
non-thermal X-ray emission. The plus sign marks the
approximate position of the pulsar. \citep[From][]{rtk+03}  }
    \label{fig:g11}
  \end{center}
\end{figure}

SNR G11.2$-$0.3 is probably the result of the historical 
guest star of 386 A.D. \citep{cs77}. The well determined age 
\citep{tr03}, highly 
spherical morphology of the SNR shell and the likelihood of nearly constant 
energy injection by the pulsar (indicated by its current long spin-down 
timescale) make it a particularly simple system to study \citep{krv+01}. 
Figure~\ref{fig:g11} shows the relationship between the thermal X-ray 
emission (red), the non-thermal X-ray emission (blue) and the radio 
emission (green). Note the separation between the shell and the 
central PWN, and the thermal emission outside the radio PWN. In this system, 
the SNR reverse shock has probably not begun to interact with the PWN.

\begin{figure}
  \begin{center}
    \includegraphics[width=\columnwidth]{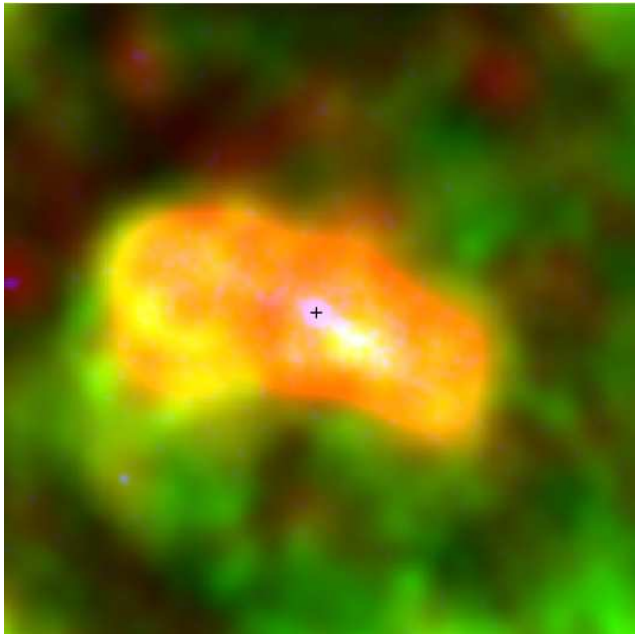}
    \caption{\small Close-up Chandra and VLA image of G11.2$-$0.3. 
The red is highly smoothed non-thermal X-rays, the green and blue as in 
Fig.~\ref{fig:g11}.}
    \label{fig:g11_pwn}
  \end{center}
\end{figure}

Figure~\ref{fig:g11_pwn} shows a close-up of the PWN, but this time the red
is a highly smoothed image of the non-thermal X-ray emission to emphasise the 
fainter emission. The X-ray nebula has bright, narrow components 
on top of a broad faint component which mostly fills the radio nebula. 
The radio appears to outline the X-ray nebula with an edge brightened appearance
except for a coincidence with the bright X-ray ``spot"  region to the
southeast of the pulsar \citep{rtk+03}. In a modified \citet{kc84} picture, 
the fresh particles may be injected in a narrow jet and remain fairly well
collimated out to the nebular edge. The region of rapid bulk motion of 
X-ray emitting particles may be at a lower pressure than the surrounding 
regions, resulting in a lower magnetic field and particle density. Hence,
the radio emission is suppressed. If there is significant cooling of 
the X-ray emitting particles in the flow, then the brightest X-ray emission
may come from the rapid flow region. It also might be the case that the 
post termination shock magnetic field never reaches equipartition and grows
out to the forward shock, which would also cause an edge brightening
effect in the radio.

\subsection{PSR B1509$-$58}

\begin{figure}
  \begin{center}
    \includegraphics[width=\columnwidth]{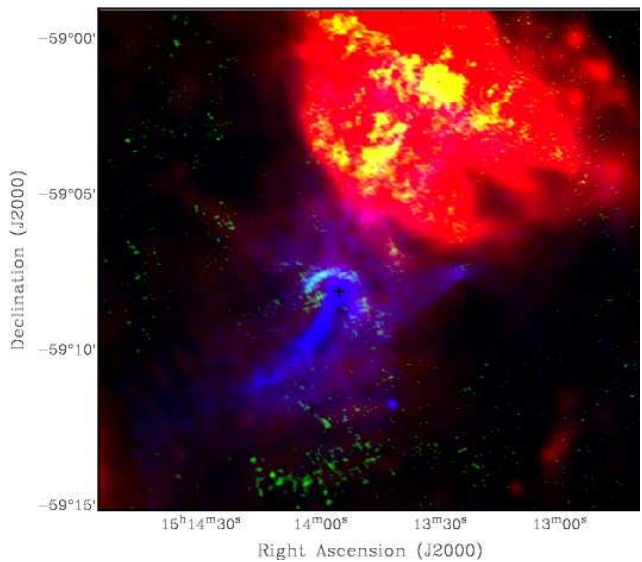}
    \caption{\small Chandra and ATCA image of the PWN around PSR B1509$-$58 
and a portion of SNR G320.4$-$1.2. Red is 20~cm radio, 
blue is non-thermal X-ray, and green is 6~cm linear polarization intensity. 
}
    \label{fig:1509_large}
  \end{center}
\end{figure}

The PWN around the $\sim 1000$~yr old pulsar PSR B1509$-$58 in the SNR 
G320.4$-$1.2 has an unusually complicated morphology. Perhaps the most 
remarkable feature of the X-ray PWN is the long, bright, jet-like flow
(see Figure~\ref{fig:1509_large}). Comparison with the radio emission shows a 
strong anti-correlation of the bright X-ray emission with the radio
\citep{gak+02}. Similar to G11.2$-$0.3, but here even more pronounced,
the radio emission seems to outline a cavity containing the brightest
X-ray emission. 

\begin{figure}
  \begin{center}
    \includegraphics[width=\columnwidth]{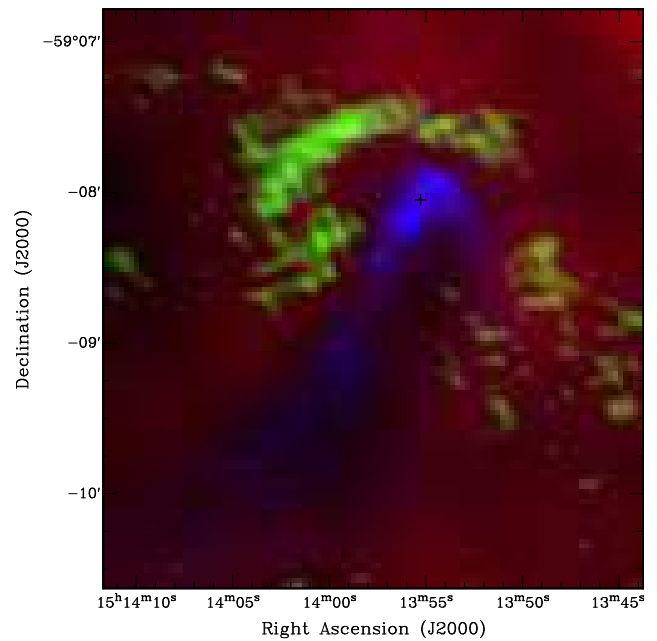}
    \caption{\small Close-up of region near PSR B1509$-$58. Colors same as in 
Fig.~\ref{fig:1509_large}.}
    \label{fig:1509_small}
  \end{center}
\end{figure}

The arc near the pulsar (Figure~\ref{fig:1509_small}) may be a toroidal 
outflow or some bow-like structure where the jet in the forward direction
suddenly ends. Note that the high linear polarization denotes a region of
highly structured magnetic field. 

\section{RPWN Associated With Gamma-Ray Sources}

\begin{figure}
  \begin{center}
    \includegraphics[width=\columnwidth]{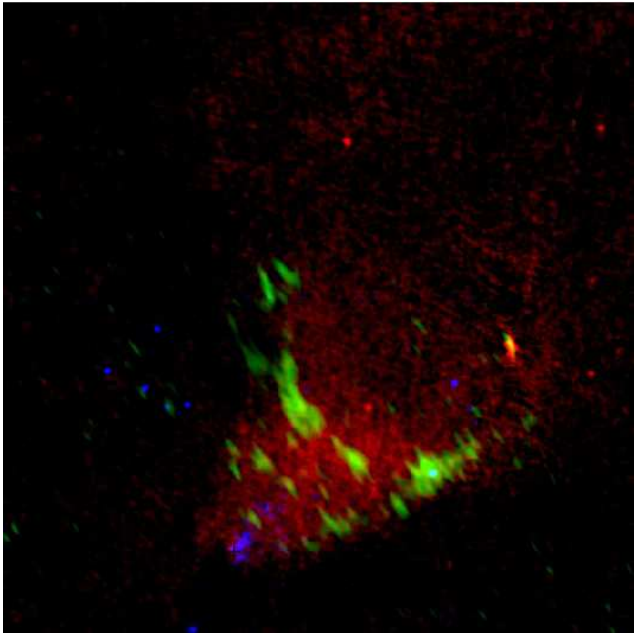}
    \caption{\small Chandra and VLA image of the nebula in the $EGRET$ source
GeV J1809$-$2327. Red is 20~cm radio continuum, green is 6~cm polarization, 
blue is non-thermal X-ray.}
    \label{fig:1809}
  \end{center}
\end{figure}

\begin{figure}
  \begin{center}
    \includegraphics[width=\columnwidth]{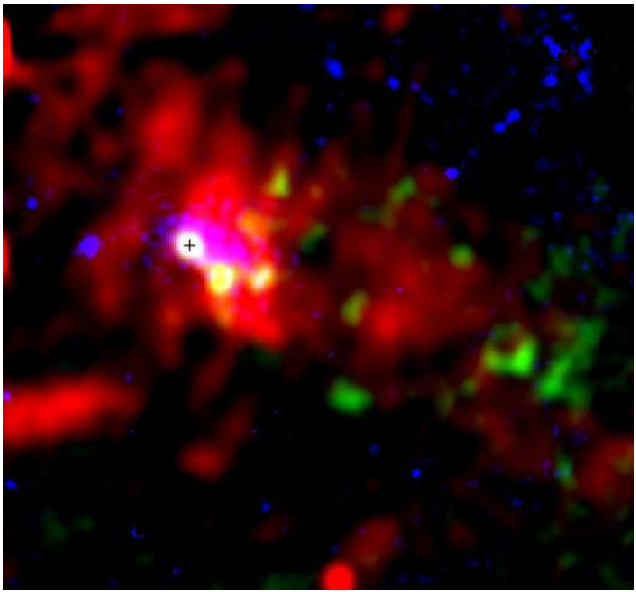}
    \caption{\small XMM-Newton and ATCA image of Rabbit nebula and lower
wing of the Kookaburra radio complex in GeV J1417$-$6100. Red is 20~cm 
radio continuum, green is 20~cm linear polarization, blue is non-thermal X-ray.}
    \label{fig:rabbit}
  \end{center}
\end{figure}

RPWN are formed after the pulsar has moved significantly from its birthplace
and are often found outside their associated SNR. Hence, they are usually
surrounded by the ISM, and their motion is probably supersonic causing
a bow-shock to form. Also, since they are on average older than sPWN, 
they tend to have lower spin-down energies. Despite this, they are often found 
to be coincident with unidentified $\gamma-$ray sources. Two apparent 
rPWN are shown in Figures~\ref{fig:1809} \citep{brrk02} and \ref{fig:rabbit}
\citep{rrjg99}. The X-ray emission again seems collimated while the radio 
emission forms a sheath around it. The polarized emission seems to outline the 
nebula (although the data isn't completely clear). More detailed
radio polarization studies could clarify the magnetic field structure. 
The current structure might hint that the radio emitting particles may
not be accelerated near the pulsar but instead at the nebular/ISM boundary.

\section*{Acknowledgments}

This work was partially supported by NASA through SAO grant GO3-4098X.


\begin{thebibliography}{}
\setlength\itemsep{0cm}

\bibitem[{Braje} {et al.}  (2002)]{brrk02}
{Braje}, T.~M., {Romani}, R.~W., {Roberts}, M.~S.~E., \& {Kawai}, N. 2002, ApJ,
  565, L91

\bibitem[Clark \& Stephenson(1977)]{cs77} Clark, D.H. \& Stephenson, F.R.
1977, {\it The Historical Supernovae}, Pergamon, Oxford.

\bibitem[Gaensler {et al.}  (2002)]{gak+02}
Gaensler, B.~M., Arons, J., Kaspi, V.~M., Pivovaroff, M.~J., Kawai, N., \&
  Tamura, K. 2002, ApJ, 569, 878

\bibitem[Kaspi et al. (2001)]{krv+01}
Kaspi, V.~M., Roberts, M. S.~E., Vasisht, G., Gotthelf, E.~V., Pivovaroff, M.,
\& Kawai, N. 2001b, ApJ, 560, 371

\bibitem[Kaspi, Roberts \& Harding(2004)]{krh04} 
Kaspi, V.~M., Roberts, M. S.~E., \& Harding, A.~K. 2004, to appear in 
{\it Compact Stellar X-ray Sources} ed. Lewin, W.H.G. \& van der Klis, M. 
astro-ph/0402136

\bibitem[Kennel \& Coroniti (1984)]{kc84}
Kennel, C.~F. \& Coroniti, F.~V. 1984, ApJ, 283, 694

\bibitem[Roberts {et al.} (1999)]{rrjg99}
Roberts, M. S.~E., Romani, R.~W., Johnston, S., \& Green, A.~J. 1999, ApJ, 515,
  712

\bibitem[Roberts et al.  (2003)]{rtk+03}
Roberts, M. S.~E., Tam, C.~R., Kaspi, V.~M., Lyutikov, M., Vasisht, G.,
Pivovaroff, M., Gotthelf, E.~V., \& Kawai, N. 2003, ApJ, 588, 992

\bibitem[{Tam} \& {Roberts}(2003)]{tr03}
{Tam}, C. \& {Roberts}, M.~S.~E. 2003, ApJ, 598, L27


\end{thebibliography}
\end{document}